\documentclass[conference]{IEEEtran}
\IEEEoverridecommandlockouts
\ifCLASSINFOpdf
 \usepackage[pdftex]{graphicx}
\else
\fi

\usepackage{multirow}
\usepackage{array,comment}
\usepackage[lofdepth,lotdepth]{subfig}

	\usepackage[utf8]{inputenc} 
	\usepackage[T1]{fontenc}

\AtBeginDocument{%
  \renewcommand\tablename{TABLO}
}

\AtBeginDocument{%
  \renewcommand\abstractname{Abstract}
}

\begin{document}

%
\title{Bipolar Bozuklukta Otomatik Mani Değerlendirmesi için Konuşma Analizi\\
Speech Analysis for Automatic Mania Assessment in Bipolar Disorder}

\author{\IEEEauthorblockN{Pınar Baki}
\IEEEauthorblockA{Computer Engineering\\
Boğaziçi University\\
Istanbul, Turkey\\
Email: pinarbaki95@gmail.com}
\and
\IEEEauthorblockN{Heysem Kaya}
\IEEEauthorblockA{Information and Computing Sciences\\
Utrecht University, Utrecht, the Netherlands,\\
Email: h.kaya@uu.nl}
\and
\IEEEauthorblockN{Elvan Çiftçi}
\IEEEauthorblockA{
Department of Psychiatry\\
University of Alberta\\
Edmonton, Canada,\\
Email: elvanlciftci@gmail.com}
\and
\IEEEauthorblockN{Hüseyin Güleç}
\IEEEauthorblockA{ Department of Psychiatry\\
Erenköy Psychiatric and Neurological \\ Diseases Training and \\ Research Hospital\\ Istanbul, Turkey\\
Email: huseyingulec@yahoo.com}
\and
\IEEEauthorblockN{Albert Ali Salah}
\IEEEauthorblockA{Information and Computing Sciences\\
Utrecht University, Utrecht, the Netherlands,\\
Email: a.a.salah@uu.nl}
}

\maketitle

\begin{ozet}
Bipolar bozukluk, manik ve depresif ataklara neden olan zihinsel bir hastalıktır. Bu çalışmada, Bipolar Disorder veri setindeki
7 farklı görev içeren kayıtlar, konuşma özellikleri kullanılarak hipomani, mani ve remisyon sınıflarına ayırıldı. Deneyler ses kayıtlarından elde edilen ayrı görev dosyalarında gerçekleştirildi. 6. ve 7. görevlerle birlikte eğitilen modelle en iyi performans elde edildi ve bu sonuç veri setinin temel sonuçlarından daha yüksek olan 0.53 UAR (unweighted average recall) sonucunu verdi.
\end{ozet}
\begin{IEEEanahtar}
Duyuşsal bilişim, paralinguistik, bipolar bozukluk.
\end{IEEEanahtar}

\begin{abstract}
Bipolar disorder is a mental disorder that causes periods of manic and depressive episodes. In this work, we classify recordings from Bipolar Disorder corpus that contain 7 different tasks, into hypomania, mania and remission classes using only speech features. We perform our experiments on splitted tasks from the interviews. Best results achieved on the model trained with 6th and 7th tasks together gives 0.53 UAR (unweighted average recall) result which is higher than the baseline results of the corpus.
\end{abstract}

%

\IEEEpubidadjcol

\section{G{\footnotesize İ}r{\footnotesize İ}ş}

Bipolar bozukluk kişinin duygu durumunda şiddetli değişimlere sebebiyet veren bir hastalıktır. Bu değişimler duygusal yükselmeler (manik dönem), alçalmalar (depresyon) veya karışık bir biçimde gözlemlenebilir. Bipolar hastalığını teşhis etmek için hastalar üzerinde uzun süreler boyunca gözlem yapmak gerekir, aksi takdirde endişe veya depresyon gibi psikolojik hastalıklarla karıştırılabilir. Günümüzde, akıl hastalıklarının teşhisi psikiyatristler tarafından yapılan testlere ve hastaların veya yakınlarının verdiği raporlara dayanmaktadır. Ancak bu raporların objektif olmamasından dolayı sistematik ve daha objektif bir metotun geliştirilmesine ihtiyaç duyulmaktadır.

Konuşmadaki değişimler psikolojik hastalıkların tespitinde önemli bir semptom olarak kabul edilmektedir. Hızlı, gürültülü veya aceleye gelmiş hissiyatı uyandıran konuşmalar, kontrol edilemeyen düşünceler bazı yaygın konuşma bozukluklarına örnek olarak verilebilir. Öte yandan, konuşma bozukluklarını tespit etmek diğer semptomlara nazaran daha kolaydır. Bu semptomların hastalar tarafından saklanması zordur ve konuşma dili farketmeksizin benzer özelliklere sahip oldukları için genelleştirilmeleri kolaydır. Konuşma verisinin elektronik aletlerle hızlı ve kolay bir şekilde toplanabilmesi de bu alanda çalışma yapmak için önemli bir motivasyon olmuştur. \cite{low2019review}.

Son yıllarda zihinsel problemlerin ve hastalıkların tespiti için yapay öğrenme teknikleri değerlendirilmeye başlandı. Bu amaçla yapılan çalışmalardan hem görsel [2], hem de işitsel kiplerin kullanıldığını görüyoruz. Genellikle en iyi sonuçlar iki ya da daha çok kip birlikte kullanıldığında elde ediliyor. Bu hastalıkların tespitinde yapay zeka sistemlerinin kullanılmasının bazı avantajları var. Hastayı evinde ve daha sık takip edebilmek, hastada gelişen davranış değişikliklerinin hızlıca yakalanması, teşhisin kolaylaşması ve semptomların uzun zaman aralıklarında izlenerek analizinin yapılması avantajlar arasında sayılabilir. \cite{low2019review} Öte yandan yapay zeka sistemlerine fazlaca güvenmek ve uzmanların teşhis yeteneklerinin buna bağlı olarak zaman içinde körelmesi ise bu yaklaşımların olası riskleridir.

Bu çalışmada bipolar hastalar için hastalık derecesinin işitsel sınıflandırmasını otomatik olarak yapacak bir sistem önerildi. Çalışma için 2018 Audio/Visual Emotion Challange (AVEC)'da kullanımı herkese açılan Bipolar Disorder veri seti kullanıldı. Makalenin 2. bölümünde bu veri setiyle yapılan diğer çalışmalardan, 3. bölümde veri setinin ayrıntılarından, 4. bölümde hazırlanan modelden, 5. bölümde deney sonuçlarından ve 6. bölümde de elde edilen sonuçlardan bahsedilmiştir.

\IEEEpubidadjcol

\section{İLGİLİ ÇALIŞMALAR}
\label{özet}
\begin{table*}[t]
\begin{center}
\caption{BD veri setini kullanan çalışmaların karşılaştırması. Skorlar makalelerde belirtilen en yüksek UAR skorlarıdır. }
\begin{tabular}{|c|c|c|c|c|}
\hline
\textbf{Makaleler}                                                                           & \textbf{Öznitelikler}                        & \textbf{Sınıflandırıcı} & \textbf{Geçerleme} & \textbf{Test} \\ \hline
Ringeval v.d. \cite{ringeval2018avec}         & eGEMAPS+FAUs  & SVMs & 0.550                & 0.500         \\ \hline
Yang v.d. \cite{yang2018bipolar}         & Arousal and upper body posture features  & Multistream & 0.783                & 0.407         \\ \hline
Du v.d. \cite{du2018bipolar} & MFCC                                     & IncepLSTM & 0.651                & -             \\ \hline
Xing v.d. \cite{xing2018multi}    & eGEMAPS+MFCC+AUs+eyesight features       & Hierarchical recall model & \textbf{0.867}                & \textbf{0.574}         \\ \hline
Syed, Sidorov, Marshall \cite{syed2018automated}                   & AUs+gaze+pose                            & GEWELMs & 0.550                & 0.482         \\ \hline
Ebrahim, Al-Ayyoub, Alsmirat \cite{ebrahim2018determine}   & MFCC+eGEMAPS+BoAW+DeepSpectrum+FAUs+BoVW & Bi-LSTM & 0.592                & 0.444         \\ \hline
Amiriparian v.d. \cite{schuller2019capsule}                   & Mel-Spectogram                           & CapsNet & 0.462                & 0.455         \\ \hline
Ren v.d. \cite{ren2019multi} & MFCC                                     & Multi-instance learning & 0.616                & \textbf{0.574}         \\ \hline
\end{tabular}
\end{center}
\end{table*}
Akıl sağlığı bozukluklarının makine öğrenimi yöntemleri kullanarak incelenmesi ve teşhis edilmesi aktif bir araştırma alanıdır. Psikiyatristler ve bilgisayar bilimcileri arasındaki disiplinler arası çalışmalar, yeni veri kümeleri oluşturmaya ve tıp alanından yapay zeka uygulamalarına yeni bakış açıları getirilmesine yardımcı olmaktadır. Oluşturulan veri setlerinin birçoğu hem görsel hem de işitsel kipleri barındırmaktadır. Bu da araştırmacıların bilgisayarla görme, konuşma işleme ve doğal dil işleme gibi farklı modeller kullanarak yöntemler geliştirmelerine olanak sağlamaktadır. En gelişmiş sonuçlar, farklı kiplerle oluşturulan modellerin birleştirilmesiyle elde edilmektedir.

Geçtiğimiz yıllarda, bipolar hastalığının otomatik tespiti konusunda birçok makale ve veri seti \cite{khorram2018priori} yayınlanmıştır. Bunlardan 2018 yılında  Audio/Visual Emotion Challenge (AVEC) \cite{ringeval2018avec} yarışmasında kullanıma sunulan Bipolar Disorder (BD)  veri seti üzerinde birçok çalışma yapılmıştır \cite{yang2018bipolar, du2018bipolar, xing2018multi, syed2018automated, ebrahim2018determine, schuller2019capsule, ren2019multi}. Bu çalışmaların birçoğunda \cite{yang2018bipolar, xing2018multi, syed2018automated, ebrahim2018determine} hem görsel hem işitsel özellikler çıkarılmış, ve karar aşamasında bu iki kipten gelen tahmin birleştirilerek veya çıkarılan özellikler karar aşamasından önce birleştirilerek sonuçlar elde edilmiştir. Bütün çalışmalarda en iyi sonuçlar bu iki kip birlikte kullanılarak elde edilmiştir. BD veri setinde 46 farklı kişiden toplamda 218 örnek bulunmaktadır. Bu yüzden, derin öğrenme modellerinde bu veri seti üzerinde çalıştırırken aşırı öğrenme problemiyle karşılaşılmaktadır. Bu problem geçerleme ve test setleri arasında önemli derecede performans düşüşüne sebep olabilmektedir.  \cite{du2018bipolar} çalışmasında bu problemi, LSTM ağında L\textsubscript{1} düzenlileştirme (regularization) yöntemini kullanarak çözmüşlerdir. Benzer şekilde \cite{ren2019multi}'de, yazarlar aşırı öğrenme problemini çözmek için, BD veri seti üzerinde derin sinir ağı eğitirken çoklu-örnekli öğrenme (multi-instance learning) yöntemi kullanmışlardır.

\section{VERİ SETİ}

Bu çalışmada, 2018 Audio/Visual Emotion Challenge (AVEC) yarışmasında yayınlanan Bipolar Disorder veri seti~\cite{cciftcci2018turkish} kullanılmiştir. BD veri setinde  bipolar bozukluğu olan 46 hastanın ve 49 sağlıklı deneğin verisi bulunmaktadır. Hastaların depresif ve manik durumları Young Mania Rating Scale (YMRS) ve Montgomery-Asberg Depression Rating Scale (MADRS) ölçütleri kullanılarak, hastanın hastaneye yatışının 0, 3, 7 ve 28. günlerinde ve 3. ayda taburcu olduktan sonra kaydedilmiştir. Bu günlerde, görsel ve işitsel kayıtlar alınmış ve kayıdın ardından gerekli etiketlemeler yapılmıştır. Etiketleme bipolar bozukluk durumuna (mani, hipomani, depresif) ve YMRS skorlarına göre yapılmıştır. Ayrıca, veri setinde hastalarla ilgili yaş ve cinsiyet gibi veriler de bulunmaktadır. 

Kayıtlarda, hastalardan yedi farklı görevi gerçekleştirmeleri istenmiştir. Görevler hastalarda farklı duygu durumlarını ortaya çıkaracak şekilde tasarlanmıştır. Bu sayede hastalar farklı şartlar altında gözlemlenebilecektir. İlk üç görev olumsuz duygu ortaya çıkaran görevler olarak kabul edilebilir, sonraki iki görev nötr, herhangi bir duygu ifade edilmesi beklenmeyen görevlerdir ve son iki görev de olumlu duygu ortaya çıkaran görevlerdir. Gerçekleştirilen görevler hastaneye gelme nedenini açıklamak, üzücü bir anıyı anlatmak, Van Gogh'un Depresyon resmini açıklamak, birden otuza kadar saymak, birden otuza kadar daha hızlı bir şekilde saymak, mutlu bir anıyı anlatmak ve Dengel'in Ev Tatlı Ev resmini açıklamaktır.

Bu çalışmada veri setindeki bipolar hastalarının kayıtlarıyla deneyler gerçekleştirilmiş ve sağlıklı deneklerin kayıtları kullanılmamıştır. Bipolar hastalarının duygu durumları hastaların doktorları tarafından mani, hipomani ve remisyon olmak üzere üç sınıfla etiketlenmiştir. AVEC yarışmasında, veri setinin eğitim, geçerleme ve test kümeleri, katılımcalara ayrı setler halinde verilmiş ve katılımcalardan sonuçlarını bu kümelemelere göre raporlamaları istenmiştir. Bu çalışmada da, hazırlanan modelle alınan sonuçları veri seti üzerinde çalışan önceki yöntemlerle karşılaştırabilmek amacıyla kümelemeler aynı şekilde kullanılmıştır. Eğitim, geçerleme ve test kümeleri sırasıyla 104, 60 ve 54 örnek içermektedir. Yarışmada sonuçları bildirmek için UAR skoru kullanılmıştır. UAR, her bir sınıfın (mani, hipomani, remisyon) duyarlılık (recall) skorunun ağırlıksız ortalaması alınarak hesaplanmaktadır.

\section{SİSTEM TANIMI}
\begin{figure*}[t]
\begin{center}
 \includegraphics[scale=.5]{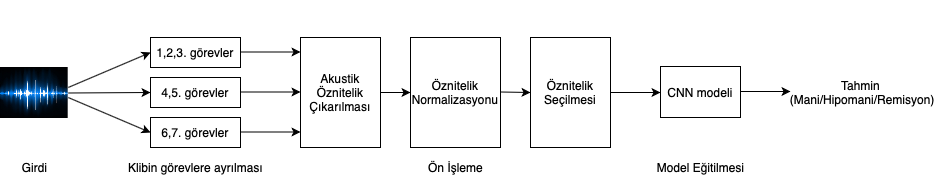}
 \caption[system]{Önerdiğimiz yöntemin şematik gösterimi.}
 \label{fig:system}
\end{center}
\end{figure*}

Geliştirilen sistem ile Bipolar Disorder veri setinde~\cite{cciftcci2018turkish} hastanın remisyon, hipomani ve mani aşamalarının hangisinde olduğu tespit edilmeye çalışıldı. Bunun için akustik öznitelik çıkarma, öznitelikleri normalize etme, seçme ve kısımları geliştirildi (bkz. Şekil~\ref{fig:system}).

\subsection{Öznitelik çıkarma}
Akustik öznitelikler açık kaynaklı openSMILE aracı kullanılarak konuşma sinyallerinden elde edilmiştir. openSMILE konuşma işleme yöntemleri için bazı örnek yapılandırma dosyaları sağlamaktadır. Bu yapılandırma dosyaları INTERSPEECH yarışmalarında kullanılan temel akustik öznitelikleri otomatik olarak oluşturmaktadır. Bu çalışmada INTERSPEECH 2010 \cite{schuller2010interspeech} temel öznitelikleri kullanıldı. Bu öznitelikler, ses dosyalarından çıkarılan düşük seviyeli tanımlayıcılara (LLD) (MFCC, skewness, kurtosis vb.), standart sapma, ortalama alma gibi bazı fonksiyonlar kullanılarak elde edilmektedir. INTERSPEECH 2010 yapılandırma dosyasıyla, bu yöntemle elde edilen 1582 akustik öznitelik çıkarılmaktadır.

\subsection{Ön İşleme}
Konuşma sinyallerinin sayısını artırmak ve yapılan farklı görevlerin etkisini incelemek için, öznitelikleri elde etmeden önce ses dosyaları görevlere ayrılarak kaydedildi. Ses kayıtlarında her görevin arasında görevin başladığını belirten bir zil sesi bulunmaktaydı. Ancak bazı kayıtlarda, hastalar bu sesten sonra da önceki görev hakkında konuşmaya devam etmiş ya da yanlışlıkla soruyu geçip geri dönmüştü. Bu da görevleri otomatik ayırırken bazı hatalara yol açtı. Bu hataları düzeltmek amacıyla yanlış işaretlenen görev başlangıç ve bitiş zamanları tekrar dinlenerek işaretlendi. Görev zamanları işaretlenip her bir görev için ses dosyaları oluşturulduktan sonra eğitim setinden openSMILE aracıyla çıkarılan özniteliklere z-normalizasyon uygulandı. Geçerleme ve test setleri de bu dağılıma uyduruldu. Kullanılan INTERSPEECH 2010 öznitelik seti 1582 akustik öznitelik içermektedir. BD veri setinin az sayıda veri içerdiği de göz önüne alındığında, bu geniş öznitelik setini kullanmak aşırı öğrenme problemini önlemeyi zorlaştırıyordu. Bu sebeple, öznitelik setinin boyutunu azaltmak için öznitelik seçimi yöntemi uygulandı. Özyinelemeli (recursive) öznitelik eleme yöntemi, model doğruluğunu kullanarak her aşamada modele en az katkısı olan özniteliği seçip eler. Bu çalışmada önyinelemeli öznitelik eleme yöntemi SVM modeliyle kullanıldı ve her bir ses dosyası için öznitelik seti boyutu 100 özniteliğe indirildi. Yapılan deneyler sonucu hem aşırı öğrenme  problemini azaltacak hem de en iyi performans alınacak öznitelik sayısı seçildi. İleriki çalışmalarda öznitelik sayısının ve farklı özniteliklerin performansa etkisi de incelenecektir.

\subsection{Model}
Şu ana kadarki çalışmada, akustik öznitelikleri 1D evrişimli sinir ağı (convolutional neural network) (CNN) modeli üzerinde eğitilerek kullanıldı. Model 1 evrişimli girdi katmanı, 25 nöron içeren 1 evrişimli gizli katman ve her sınıf için birer nöron içeren 3 nöronlu bir çıktı katmanından oluşmaktadır. Aşırı öğrenme problemini engellemek için, her katmandan sonra batch normalization ve dropout katmanları eklendi ve erken sonlandırma yöntemi kullanıldı. Geçerleme yitimi 5 kere artarsa, eğitim sonlandırıldı. Bu yolla modelin eğitim setini ezberlemesi engellendi.

\section{DENEY SONUÇLARI}
\begin{figure}[t]
    \begin{centering}
        \includegraphics[width = 0.7\linewidth]{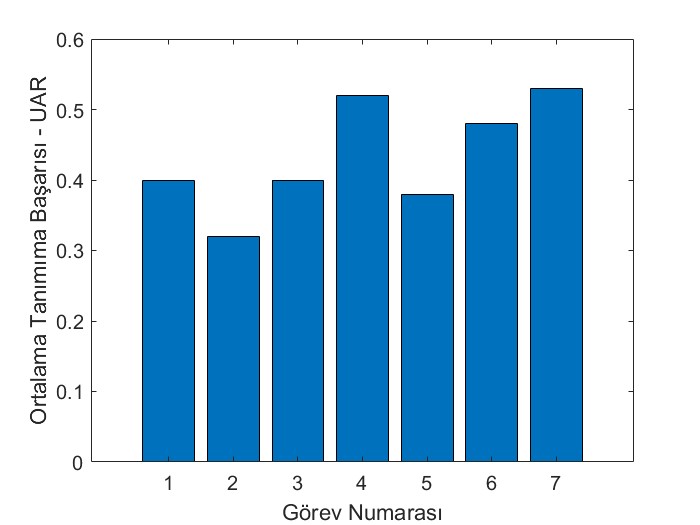}
\caption{Farklı görevlerdeki veriyle elde edilen UAR sonuçları. Görevlerde anlatılanlar sırasıyla: 1-Hastaneye geliş sebebi, 2-Van Gogh'un Depresyon tablosunun hissettirdikleri, 3-Kötü bir anı, 4-Birden otuza kadar sayma, 5-Birden otuza kadar hızlı bir şekilde sayma, 6-Dengel'in Home Sweet Home tablosunun hissettirdikleri, 7-Mutlu bir anı.}
\label{figure:taskResults}   
\end{centering}
\end{figure}

İlk deney görevlere ayrılmış ses dosyalarında gerçekleştirildi. Farklı görevlerin etkisini gözlemlemek için model bütün görevlerle ayrı ayrı eğitildi. Şekil \ref{figure:taskResults} her görevin UAR skorunu göstermektedir.  Mutlu duyguları harekete geçiren görevlerin (6. ve 7. görevler) ve 30'a kadar sayma görevinin (4.görev) diğer sonuçlara göre daha iyi sonuç verdiği gözlemlendi. Öte yandan, bazı hastaların kötü anıları hakkında detaylı konuşmak istememesi ve/veya kısa cevaplar vermelerinin bazı görevlerdeki (1.,2., ve 3.) düşük performasın altında yatan sebep olduğunu düşünüyoruz. Son olarak, 5.görevdeki performans düşüklüğünün sebebinin hızlı konuşmanın, konuşmadaki ayırt edici özelliği ortadan kaldırmasından dolayı olduğunu düşünüyoruz. Resim \ref{fig:7} yapılan deneylerin 7.görevde eğitilen modelde ve 7. görevlere oluşturulan geçerleme setindeki sonuçları gösteriyor. Üç sınıfın da şans seviyesinden daha yüksek performans gösterdiğini görüyoruz (0.33).

\begin{figure}[t]
    \begin{centering}
        \includegraphics[width = 0.7\linewidth]{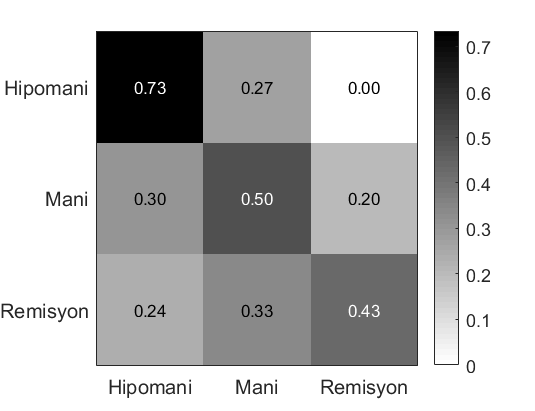}
        \caption{En iyi UAR değerini veren 7. görevle eğitilen modelin geçerleme setiyle oluşturulan hata dizeyi}
        \label{fig:7}
    \end{centering}
\end{figure} 

Sonuçların verisetini kullanan diğer çalışmalarla da karşılaştırılabilmesi amacıyla, 7. görev kullanılarak eğitilen ve en iyi sonucu veren model tüm ses kaydından çıkarılan görevlerle oluşturulan geçerleme seti üzerinde de test edildi. Orjinal geçerleme setindeki dosyalardan INTERSPEECH 2010 temel öznitelikleri elde edildikten sonra, bu öznitelikler eğitim setini normalize ederken kullanılan dağılıma uyduruldu. Aynı zamanda, eğitim setinde seçilen en önemli 100 öznitelik geçerleme setinde de seçildi. 7. görevlerden oluşan eğitim setiyle eğitip, orjinal geçerleme setiyle test edilen modelden 0.51 UAR skoru alındı. Bu sonuç da \cite{cciftcci2018turkish} 'da gösterilen temel sonucunun üstündedir.

\begin{figure}[t]
    \begin{centering}
        \includegraphics[width = 0.7\linewidth]{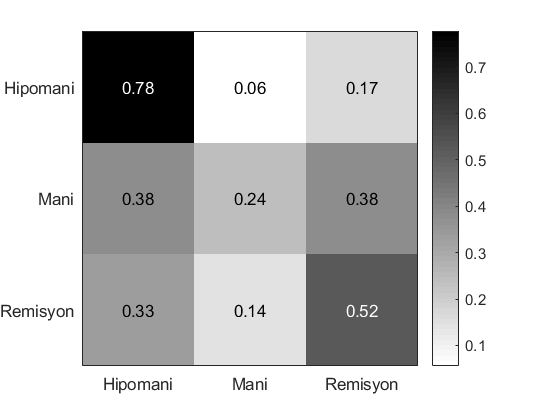}
        \caption{7. görevde eğitilip geçerleme setinin görevlere ayrılmamış versiyonunda test edilen modelin hata dizeyi}
        \label{fig:7}
    \end{centering}
\end{figure} 

Farklı görev gruplarının sınıflandırma sonuçları üzerindeki etkisini de incelemek istedik. 1,2 ve 3. görevler mutsuz duyguları açığa çıkaran görevler oldukları için negatif görevler, 4 ve 5. sayma görevleri nötr görevler, 6 ve 7. görevleri de mutluluk uyandıran duygular oldukları için pozitif görevler olarak gruplandırıldı. Önceki deneylere (bkz. Şekil \ref{figure:taskResults}) benzer şekilde en iyi sonuç 6. ve 7. görev birlikte kullanılarak eğitilen modelde elde edildi. 

\begin{table}[h]
\caption{Farklı görev grupları kullanılarak yapılan deneylerde elde edilen UAR skorları. İlk sütun eğitim yapılırken hangi görevlerin birlikte gruplandırıldığını, ikinci sütun eğitilirken kullanılan görev gruplarını geçerleme setinde de kullanıldığında elde edilen UAR skorlarını, üçüncü sütun da geçerleme sırasında bütün kayıt dosyası kullanılarak elde edilen UAR skorlarını göstermektedir.}
\label{table:taskGroups}   
\begin{center}
\begin{tabular}{|c|c|c|}
\hline
\textbf{Görev No} & \textbf{Seçili Görevlerdeki UAR} & \multicolumn{1}{l|}{\textbf{Tüm Kayıttaki UAR}} \\ \hline
1-2-3                & 0.46                           & 0.47                                                     \\ \hline
4-5                  & 0.34                           & 0.36                                                     \\ \hline
6-7                  & 0.46                           & 0.53                                                     \\ \hline
\end{tabular}
\end{center}
\end{table}

\section{ANALİZ VE VARGILAR}
Bu makalede Bipolar Disorder verisetinde elde ettiğimiz sonuçları gösterdik. Yalnızca konuşma kipini kullanarak üç sınıflı (remisyon, hipomani, mani) sınıflandırma gerçekleştirdik. Öznitelik çıkarmak için openSMILE aracındaki INTERSPEECH 2010 temel öznitelik setini kullandık. Ön işleme aşamasında, çıkardığımız özniteliklere normalizasyon ve öznitelik eleme yöntemlerini uyguladık. En iyi sonuçları 6. ve 7. görevleri (mutlu hisler uyandıran bir tabloyu anlatma ve mutlu bir anı anlatma) birlikte kullanarak eğittiğimiz modelde elde ettik. Elde edilen UAR skoru \cite{cciftcci2018turkish} gösterilen temel sonucun üstündedir. Bu çalışma, Bipolar Disorder veri setindeki görevlerin sınıflandırmaya etkisini inceleyen ilk çalışmadır. Sadece ses kipi kullanılmasına rağmen temel sonucun üstünde bir başarı elde edilmiştir.  İlerideki çalışmalarımızda, farklı görevlerin etkilerini yazımsal ve görsel kiplerde de incelemeyi ve bu kipleri alternatif yöntemlerle tümleştirerek başarımı artırmayı hedefliyoruz.

%

\bibliographystyle{IEEEtran}

\bibliography{references}

%



\end{document}